\documentclass[10pt]{amsart}
\usepackage{amsmath, amsfonts, amsthm, amssymb}
\usepackage[a4paper, width=16cm, top=30mm, bottom=30mm]{geometry}
\usepackage{mathrsfs}
\usepackage{stmaryrd}
\usepackage{verbatim}
\usepackage{hyperref}
\usepackage{color}
\usepackage{blindtext}
\usepackage{bbm}
\usepackage{ulem}
\usepackage{graphicx}
\usepackage{enumerate}
\numberwithin{equation}{section}
\newtheorem{theorem}{Theorem}[section]

\theoremstyle{definition}
\newtheorem{definition}[theorem]{Definition}
\newtheorem{remark}[theorem]{Remark}

\newcommand{\RR}{\mathbb{R}}

\newcommand{\II}{\mathbb{I}}

\newcommand{\NN}{\mathbb{N}}

\newcommand{\D}{\mathrm{d}}

\newcommand{\xx}{\mathrm{x}}

\newcommand{\LTwo}{\mathrm{L}^2}
\newcommand{\Ff}{\mathscr{F}}

\newcommand{\be}{\begin{equation}}
\newcommand{\ee}{\end{equation}}

\begin{document}

\title{Calibrating rough volatility models: a convolutional neural network approach}

\author{Henry Stone}
\address{Department of Mathematics, Imperial College London}
\email{henry.stone15@imperial.ac.uk}

\date{\today}
\thanks{We would like to thank Drew Mann, Mikko Pakkanen, Antoine Jacquier, Aitor Muguruza, Chloe Lacombe, and Blanka Horvath for feedback and helpful discussions. We also thank the EPSRC CDT in Financial Computing and Analytics for financial support.}

\maketitle
\begin{abstract}
In this paper we use convolutional neural networks to find the H\"older exponent of simulated sample paths of the rBergomi model, a recently proposed stock price model used in mathematical finance. We contextualise this as a calibration problem, thereby providing a very practical and useful application.
\end{abstract}

\section{Introduction}

The aim of this paper is to investigate whether a convolutional neural network can learn the H\"older exponent of a stochastic process, from a set of sample paths.
Recall that a stochastic process $(X_t)_{t\ge 0}$ is H\"older continuous, with H\"older exponent $\gamma>0$, if
$\vert X_t - X_s \vert \le C \vert t - s \vert^\gamma$ for  all $s,t\ge0$ and some  constant $C$. Smaller values of the H\"older exponent $\gamma$ correspond to ``rougher'' sample paths.
We shall use the terms H\"older continuity and H\"older regularity interchangeably.
 
Convolutional neural networks, referred to from here as CNNs, are known to be very powerful machine learning tools with a vast array of applications including (but of course not limited to) 
image classification \cite{HKS12}, \cite{BCGL97}, \cite{SZ15}; speech recognition \cite{DHK13}, \cite{ADY13}; and self-driving cars \cite{CKSX15}, \cite{IJKW17}. Very recently Bayer and Stemper \cite{BS18} used neural networks to learn implied volatility surfaces; the network is then used as part of a wider calibration scheme for options pricing.
To the best of our knowledge, however, this paper is the first to explore the use of CNNs to predict the H\"older exponent of a given stochastic process.

There has been a resurgent interest in fractional Brownian motion and related processes within the mathematical finance community in recent years. 
In particular, \cite{GJR18} carried out an empirical study that suggests that the log volatility behaves at short time scales in a manner similar to a fractional Brownian motion, with Hurst parameter $H \approx 0.1$.
This finding is corroborated by \cite{BLP16}, who study over a thousand individual US equities and find that the Hurst parameter~$H$ lies in $(0, 1/2)$ for each equity.
Both \cite{GJR18} and \cite{BLP16} use least square regression techniques to estimate the value of $H$. 
In this paper we instead use  a CNN to find the value of the H\"older exponent from simulated sample paths of the rBergomi model, with varying values for the H\"older exponent.
The rBergomi model, which is introduced in Subsection \ref{sec: rBergomi model}, has similar H\"older regularity properties as the fractional Brownian motion \cite[Proposition 2.2]{JPS18}.

The structure of the paper is the following. In Section \ref{intro to neural nets} we give a brief introduction to neural networks and the fractional Brownian motion.
The methodology used in the paper is outlined in Section \ref{methodology}.   
In Section \ref{solving class and reg} we use CNNs to solve the regression problem of predicting the H\"older exponent, as a continuous value, for a given sample path.
We hope to establish a robust means for calibrating rough volatility models; indeed, once the CNN has been trained we want it to perform well when making predictions on unseen data.
Thus in Section \ref{calibration section} we use the trained CNN to predict the value of the H\"older exponent on realised volatility data from financial markets; this provides a simple and accurate means of calibration.
In Section \ref{conclusion} we discuss our results and conclude the paper.
\section{An introduction to neural networks and fractional Brownian motion}\label{intro to neural nets}
We begin with an introduction to neural networks, using \cite{KNTY18} as our guide.
\subsection{Neural Networks}
An artificial neural network is a biologically inspired system of interconnected processing units, where each processing unit is called a layer. Inputs to each layer, apart from the first layer, are outputs from the previous layer.
A layer is composed of a number of nodes, and each node in a given layer is connected to the nodes in a subsequent layer, thus forming a network; each edge in this network has a weight associated to it. 
The first processing unit is called the input layer, and the final processing unit is the output layer. The processing unit or units between the input layer and output layer are referred to as hidden layers; typically artificial neural networks have more than one hidden layer. Figure \ref{NN example} below illustrates the structure of a simple artificial neural network, drawn using Python, using code available on 
Github\footnote{https://gist.github.com/craffel/2d727968c3aaebd10359}.
A formal, mathematical definition \cite[Definition 4.1]{BGTW18} of a neural network is given below in Definition \ref{neural net def}.

\begin{figure}[h!]
\centering
\includegraphics[scale=0.6]{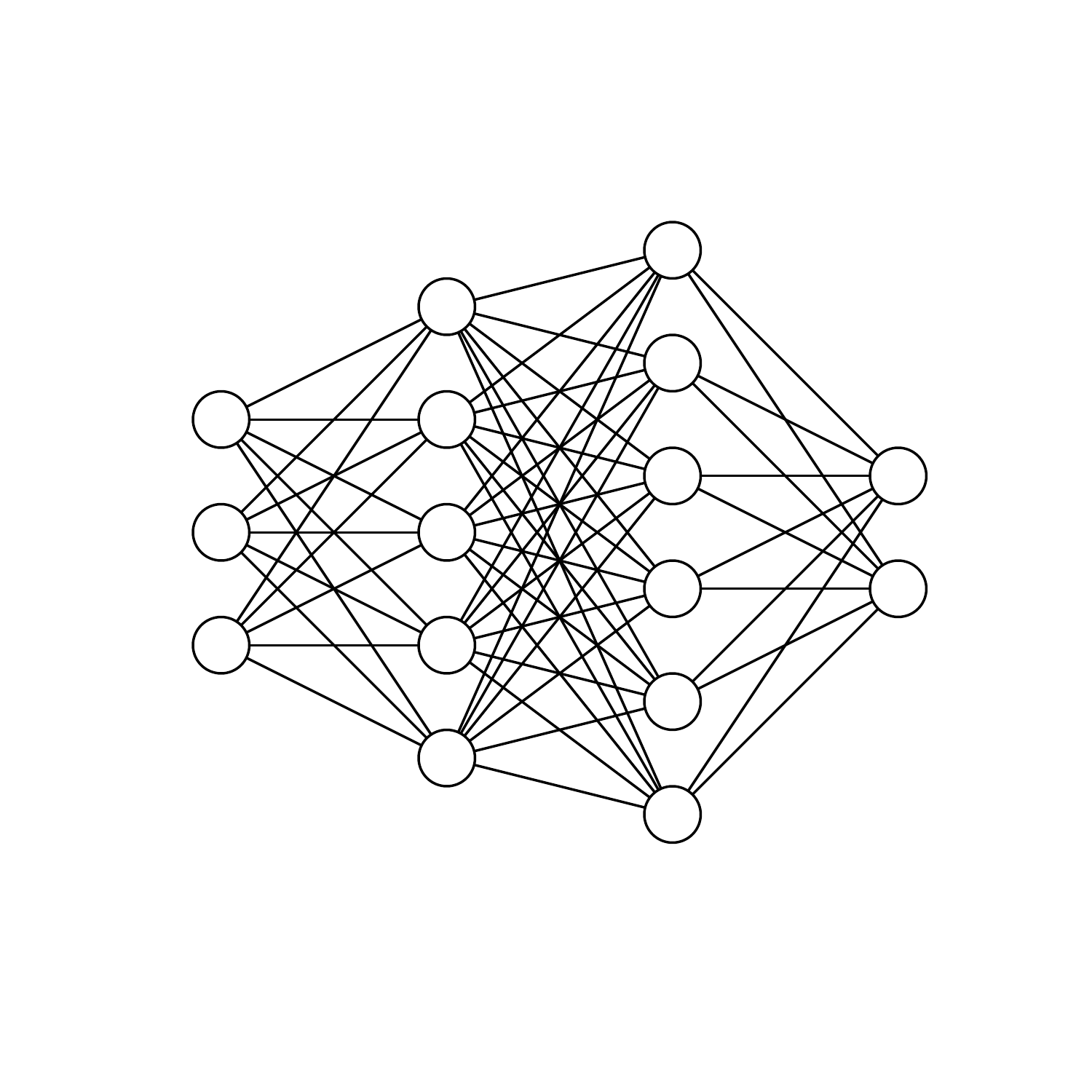}
\caption{An example of a neural network, with two hidden layers. The input layer has three nodes; the hidden layers have five and six nodes respectively; the output layer has two nodes. 
}
\label{NN example}
\end{figure}

\begin{definition}\label{neural net def}
Let $L\in\NN$ denote the number of layers in the neural network. The dimension of each hidden layer is denoted by $N_1,...,N_{L-1}\in\NN$, and the respective dimension of the input and output layer is denoted by $N_0,N_L\in\NN$.
For $A^\ell \in \RR^{N_{\ell-1}\times N_\ell} $ and $b\in \RR^{N_\ell}$
let the affine function $W_\ell:\RR^{N_{\ell-1}}\rightarrow \RR^{N_\ell} $ be defined as $W_\ell(\xx):=A^\ell \xx + b^\ell$, for $\ell=1,...,L$. The entries of matrix $A^\ell$ are the weights connecting each node in layer $\ell-1$ to layer $\ell$.
The neural network, with activation function $\sigma$, is then the function $\mathscr{N}:\RR^{N_0}\rightarrow\RR^{N_L}$ defined as the composition
\begin{equation}\label{eqn: NN}
\mathscr{N}(\xx):= W_L \circ (\sigma \circ W_{L-1})\circ ... \circ (\sigma \circ W_1) (\xx).
\end{equation}
\end{definition}

The learning process, also referred to as training, of an artificial neural network essentially boils down to finding the optimal weights in each matrix $A^\ell$ that minimise a given loss function, which depends on the task at hand: in our case, solving a regression problem. 
These optimal weights can then be used for predictions on the test set.

CNNs are a class of artificial neural networks, where the hidden layers can be grouped into different classes according to their purpose; one such class of hidden layer is the eponymous convolutional layer. Below we describe the classes of hidden layers used in our CNN. Of course, this list is not exhaustive, and there exist many classes of hidden layers that we do not describe for means of brevity. 
Note also that we describe a CNN in the context of the problem we are trying to solve, where the input data are one-dimensional vectors. CNNs can of course also be used on higher dimensional input data, but the fundamental structure and different roles of the hidden layers do not change.

\begin{itemize}
\item \textbf{Convolutional Layer:} In deep learning, the convolution operation is a method used to assign relative value to entries of input data, in our case one-dimensional vectors of time series data, while simultaneously preserving spatial relationships between individual entries of input data.
For a given kernel size $k$ and an input vector of length $m$, the convolution operation takes entries $1,...,k$ of the input vector and multiplies by the kernel element-wise, whose length is $k$. The sum of the entries of the resulting vector are then the first entry of the feature map. This operation is iterated $m+1-k$ times, thus incorporating every entry in the data vector into the convolution operation. The output of the convolutional layer is called the feature map.

For example, let $ (1,2,1,0,0,3) $ be our input vector, and $(1,0,1)$ be our kernel; here the kernel size is 3. The first iteration of the convolution operation involves taking the element-wise multiple of $(1,2,1)$ and $(1,0,1)$: 
the resulting multiple is $(1,0,1)$ and the corresponding sum of entries is 2. This is therefore the first entry of the feature map. The resulting feature map in this example is then $(2,2,1,3)$.

Clearly, the centre of each kernel cannot overlap with the first and final entry of the input vector. Zero-padding, sometimes referred to as same-padding, preserves the dimensions of input vectors and allows more layers to be applied in the CNN: zero-padding is simply the extension of the input vector and the setting of the first and final entries $1$ and to be zero, while leaving the other entries unchanged. 
In our example, the input vector becomes $ (0,1,2,1,0,0,3,0) $ after zero padding. 

\item \textbf{Activation Layer:}
The activation layer is a non-linear function that is applied to the output of the convolutional layer i.e. the feature map; the purpose of the activation layer is indeed to introduce non-linearity into the CNN. 
Such functions are called activation functions; examples include the sigmoid function and the hyperbolic tangent function.
In our CNN we use the `LeakyReLU' activation function, defined as
\begin{equation*}
f_\alpha(x) :=
\begin{cases}
\displaystyle x, & \text{if }x > 0 ,\\
\alpha x , & \text{otherwise.} 
\end{cases}
\end{equation*}
The LeakyReLU activation allows a small positive gradient when the unit is inactive. 

\item \textbf{Max Pooling Layer:} 
For a given pooling size $p$, the max pooling layer returns a vector whose entries are the maximum among the neighbouring $p$ entries in the feature map. For example, for feature map $(1,3,8,2,1,0,0,4,6,1)$ and $p=3$ the max pooling output is $(8,8,8,8,8,8,2,4,6,6)$.

Other pooling techniques apply the same idea, but use different functions to evaluate the neighbouring $p$ entries in the feature map. Examples include average pooling, and L2-norm pooling, which in fact uses the Euclidean norm in mathematical nomenclature.

\item \textbf{Dropout Layer:}
Dropout is a well-known technique incorporated into CNNs in order to prevent overfitting. 
Without the addition of a dropout layer, each node in a given layer is connected to each node in the subsequent layer; dropout temporarily removes nodes from different layers in the network. 
The removal of nodes is random and is determined by the dropout rate $d$, which gives the fraction of nodes to be temporarily dropped. 
Note that dropout is only implemented during training; during testing the weights of each node are multiplied by the dropout rate $d$.

An excellent overview of the technique is given by Hinton, Krizhevsky, Salakhutdinov, Srivastava, and Sutskever   \cite{HKSSS14}. The authors provide an extensive study to show how predictive performance of CNNs, in a number of different settings, is improved using dropout.

\item \textbf{Dense Layer:}
Also referred to as the fully connected layer, each node in the input layer is connected to each node in the output layer as the name suggests. 
After being processed by the convolutional, activation, pooling, and dropout layers, the extracted features are then mapped to the final outputs via a subset of the dense layer, an activation function is then applied subsequently.
\end{itemize}

\begin{remark}
The number of filters, the kernel size $k$, the pooling size $p$, and the dropout rate $d$ are all examples of CNN hyperparameters. 
\end{remark}

Having described the structure, we now focus on the mechanics of training the CNN. As mentioned previously, training a CNN corresponds to finding weights in the fully connected layer and kernels in the convolutional layers that minimise a specific loss function. 
Forward propagation is the name for the process by which input data is translated to an output through layers of the CNN;  it is used to give the value of the loss function, and therefore the predictive power of the CNN, for certain weights and kernels. 
The back-propagation algorithm is used to compute the gradient of the loss function from the error values of the loss function computed via forward propagation; weights and kernels, depending on the values of the loss function, are then updated iteratively. In the case of our CNN the Adam optimizer is used. 
More details on the back-propagation algorithm and the Adam optimizer can be found in \cite[Section 6.5, pages 200-219]{BCG16} and \cite{BK17} respectively.

\subsection{Fractional Brownian Motion}
The fractional Brownian motion, from here referred to as fBm, is a centred Gaussian process whose covariance function depends on a parameter $H \in (0,1)$, called the Hurst parameter. The precise definition is given below, \cite[Definition 1.1.1.]{BHOZ08}; note that setting $H=1/2$ recovers the standard Brownian motion.

\begin{definition} 
Let $H \in (0,1)$: a fractional Brownian motion $\left( W^H_t \right)_{t \ge 0}$ is a continuous, centred Gaussian process with the following covariance function
$$ \mathbb{E} \left[ W^H_t W^H_s \right] = \frac{1}{2} \left( \vert t \vert^{2H} + \vert s \vert^{2H} - \vert t-s \vert^{2H} \right), \quad \textrm{for all } s,t \ge 0.
$$
\end{definition} 

The value of the Hurst parameter $H$ completely determines the sample path H\"older regularity of a fBm. Indeed, $W^H$ has a version whose sample paths are almost surely H\"older continuous, 
with H\"older exponent $\gamma$, for all $\gamma \in (0,H)$. 

The value of $H$ also determines how the increments of $W^H$ are correlated: for $H > 1/2$, the increments of $W^H$ are positively correlated, and in this case $W^H$ is said to be persistent;
for $H< 1/2$, the increments of $W^H$ are negatively correlated, and in this case $W^H$ is said to be antipersistent.
Recall that the standard Brownian motion, in the case where $H=1/2$, has independent increments.
More details and proofs of the above statements can be found in~\cite[Chapter 1]{BHOZ08}. 

To visualise how the value of $H$ affects these characteristics we plot two sample paths of a fBm in Figure \ref{fBm plots} below. Clearly the left plot, where $H=0.1$, has a much ``rougher'' sample path than the right plot; this corresponds to a lower H\"older regularity.
The apparent mean-reversion of the left plot, compared to the 
trend-following right plot, illustrates the antipersistence and persistence of $W^{0.1}$ and $W^{0.9}$ respectively.

\begin{figure}[h!]
\includegraphics[scale=0.6]{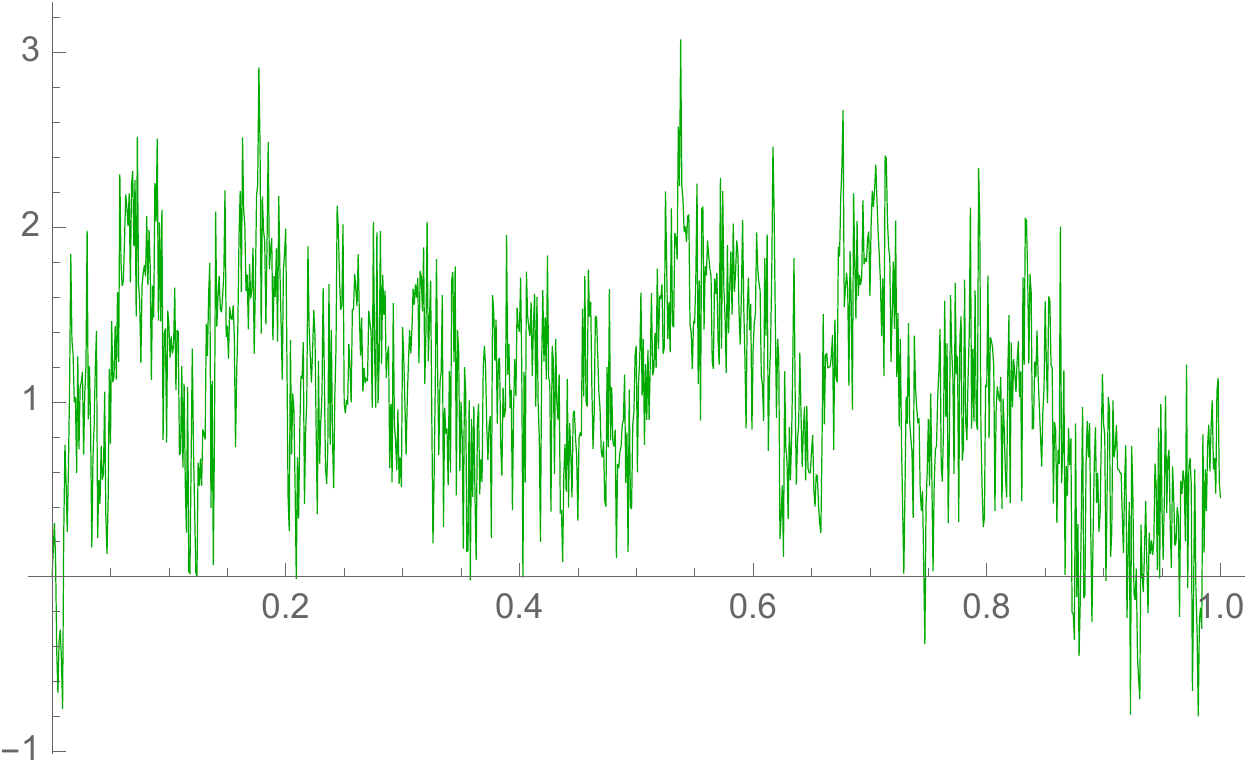}
\hspace{3mm}
\includegraphics[scale=0.6]{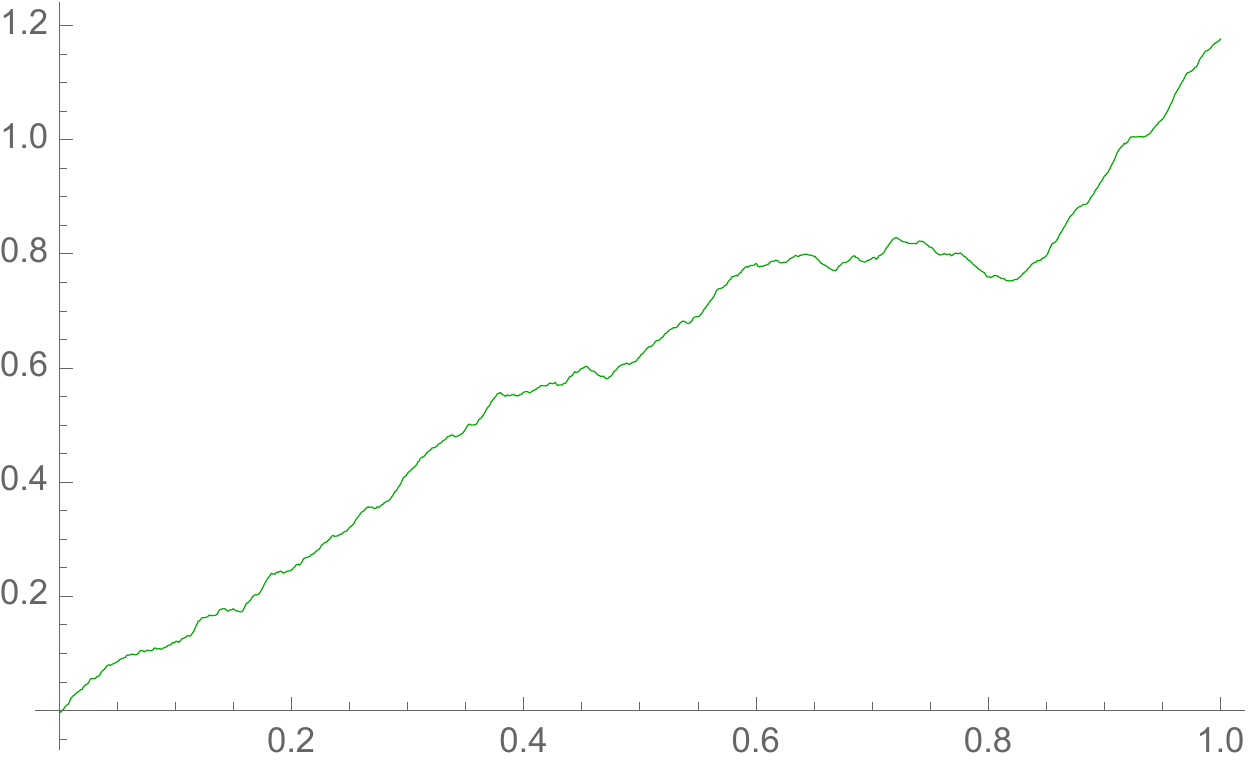}
\caption{Two sample paths of a fractional Brownian motion $W^H$ from $t=0$ to $t=1$ with 1001 sample points. In the left plot we set $H=0.1$ so that $W^H$ is antipersistent; in the right plot we set $H=0.9$, hence $W^H$ is persistent. 
Here we have used Mathematica's 
\texttt{`FractionalBrownianMotionProcess'} function to simulate each sample path.}\label{fBm plots}
\end{figure}

\begin{remark}\label{fbm integral representation remark}
Many different integral representations of fBm exist. Arguably the most well known is the characterisation by Mandelbrot and Van Ness~\cite[Definition 2.1]{MN68} as
\begin{equation}\label{Mandelbrot Van Ness Rep}
W^H_t = \frac{1}{\Gamma(H+1/2)} 
\left( \int_{- \infty }^0 ( (t-s)^{H-1/2} - (-s)^{H-1/2} ) \D B_s
+ \int_0^t (t-s)^{H-1/2}\D B_s \right),
\end{equation}
where~$B$ is a standard Brownian motion, and~$\Gamma$ the standard Gamma function.
\end{remark}

\section{Methodology}\label{methodology}
We begin this section by introducing the rBergomi model, before moving on to describe the training and test data, and the CNN architecture.

\subsection{The rBergomi Model}\label{sec: rBergomi model}
Bayer, Friz and Gatheral~\cite{BFG16} introduce a non-Markovian generalisation of Bergomi's 
`second generation' stochastic volatility model, 
 which they dub the `rBergomi' model. 
Let~$Z$ be the process defined pathwise as 
\begin{equation}\label{eq:SDEZ}
Z_t := \int_0^t K_\alpha(s,t)\D W_s,
\qquad\text{for any }t\geq 0,
\end{equation}
where $\alpha \in \left(-\frac{1}{2},0\right]$, $W$ a standard Brownian motion, 
and where the kernel 
$K_{\alpha}:\RR_+\times\RR_+ \to \RR_+$ reads
\begin{equation}\label{eq:K}
K_{\alpha}(s,t) := \eta \sqrt{2\alpha + 1}(t-s)^{\alpha}, 
\qquad \text{for all } 0\leq s<t,
\end{equation}
for some strictly positive constant~$\eta$.
Note that, for any $t\geq 0$, the map $s\mapsto K_\alpha(s,t)$ belongs to~$\LTwo$,
so that the stochastic integral~\eqref{eq:SDEZ} is well defined.
The rBergomi model, where $X$ is the log stock price process and $v$ is the variance process, is then defined as
\begin{equation}\label{rBergomi}
\begin{array}{rll}
X_t & = \displaystyle - \frac{1}{2} \int_0^t v_s \D s + \int_0^t \sqrt{ v_s } \D B_s,
 \quad &  X_0 = 0 , \\
v_t &= \displaystyle v_0
\exp \left( Z_t -\frac{\eta^2}{2}t^{2 \alpha +1}  \right),
\quad & v_0 > 0,
\end{array}
\end{equation}
where the Brownian motion~$B$ is defined as $B := \rho W + \sqrt{1-\rho^2}W^\perp$ for $\rho \in [-1,1]$
and some standard Brownian motion~$W^\perp$ independent of $W$. 
The filtration~$(\Ff_t)_{t\geq 0}$ can here be taken as the one generated by the two-dimensional Brownian motion 
$(W,W^\perp)$.

\begin{remark}
The process $\log v$ has a modification whose sample paths are almost surely locally $\gamma$-H\"older continuous,
for all $\gamma \in \left(0, \alpha + \frac{1}{2} \right)$ \cite[Proposition 2.2]{JPS18}.
As stated above, the fBm has sample paths that are $\gamma$-H\"older continuous 
for any $\gamma \in (0,H)$~\cite[Theorem 1.6.1]{BHOZ08}, so that
the rBergomi model also captures this ``roughness'' by identification $\alpha= H - 1/2$.
\end{remark}

\subsection{Training and Test Set}
We use simulated sample paths of the normalised log volatility process $( \log (v_t/v_0))_{t\ge 0}$ of the rBergomi model as our input data; with the corresponding H\"older regularity $H=\alpha+1/2 $ as the output data.
To simulate rBergomi sample paths we use Cholesky decomposition; this very well-known simulation technique is recommended because the resulting sample paths have the exact distribution, rather than an approximate distribution, of the normalised log volatility process of the rBergomi model.
The code used is publicly available on Github\footnote{https://github.com/amuguruza/RoughFCLT/blob/master/rDonsker.ipynb}.

By \cite[Proposition 2.2]{JPS18}, the H\"older regularity of the normalised log volatility process is independent of the value of $\eta$; the same proposition proves that the process $Z$, defined in \eqref{eq:SDEZ}, and the normalised log volatility process have the same H\"older regularity.  
We therefore set $\eta = 1$ in the model \eqref{rBergomi} above when generating the sample paths in Subsection \ref{rBergomi regression}, for simplicity, and we also ignore the deterministic drift term $t^{2\alpha + 1 }$. 
In Subsection \ref{robustness test regression} we take $\eta \neq 1$, and verify that this does not affect the predictive power of the CNN.

Every member of the resulting input data set has the following form: a vector $\mathbf{x}_i$, which is a rBergomi sample path with a given $H=\alpha+1/2$, and a label $\mathbf{y}_i $, which corresponds to that given $H$ used to generate $\mathbf{x}_i$.
For each $H$ value we generate 5,000 rBergomi sample paths.
We then split the input data into training and test sets; we subsequently create a validation set from part of the test set. The sizes of each training/test/validation set of the rBergomi data
are given in Table \ref{rBergomi data set sizes}.

\begin{table}[h!]
\centering
\begin{tabular}{|c|c|}
\hline
Data set & Number of samples \\
\hline
Training set & 14,000 \\
\hline
Test set & 7,500 \\
\hline
Validation set & 3,500 \\
\hline 
\end{tabular}
\caption{rBergomi input data size description.}
\label{rBergomi data set sizes}
\end{table}

\subsubsection{Selection of $H$ values}
We begin by letting $H$, and hence the corresponding label $\mathbf{y}_i$, take values in the discrete grid 
$ \{0.1,0.2,0.3,0.4,0.5 \} $.
We also sample 5 $H$ values from two probability distributions: the Uniform distribution\footnote{The corresponding set of possible $H$ values is $\{0.05, 0.18, 0.29, 0.31,0.44  \}$. }  on (0,0.5)
and the $\text{Beta}(1,9)$ distribution\footnote{The corresponding set of possible $H$ values is 
$\{0.02, 0.07, 0.06, 0.13, 0.22 \}$. }.
This not only allows us to avoid uniformity in the output of the network; it should also make the network more robust when it comes to calibration, as $H$ values for historical volatility data will almost certainly not be on the discrete grid $ \{0.1,0.2,0.3,0.4,0.5 \} $. 
Furthermore, we are also able to add emphasis to the ``rough'' values of $H$, i.e. $H \approx 0.1$, particularly in the case of the Beta distribution.

The probability density function of the $\text{Beta}(\alpha,\beta)$ distribution is given by
$ f_{\alpha, \beta}(x) = \frac{x^{\alpha-1}(1-x)^{\beta-1} }{\mathrm{B}(\alpha,\beta)}\II_{(0,1)}(x),
$ 
where the function $\mathrm{B}$ is defined as $ \mathrm{B}(\alpha,\beta):=\frac{\Gamma(\alpha)\Gamma(\beta)}{\Gamma(\alpha+\beta)}$ and $\Gamma$ is the standard Gamma function. 
We set $\alpha=1, \beta=9$, so that the expected value of the Beta distribution is 0.1, in accordance with the existing empirical studies \cite{GJR18} and \cite{BLP16}.
For each sampling method for $H$, we generate 5,000 rBergomi sample paths for each of the five $H$ values.

\subsection{CNN architecture}
We use a one-dimensional CNN, since our input $\textbf{x}_i$ are (one-dimensional) vectors, with three layers of kernels, where the kernel size for each kernel is 20 and each layer is succeeded by the Leaky ReLU activation function with alpha~$=0.1$; we add max pooling layers, each of size 3, and dropout layers between each layer of kernels. 
We use zero-padding in each of the convolutional and max pooling layers.
The values for kernel size, max pooling size, dropout rate, and rate for the Leaky ReLU activation function were chosen because they achieved the lowest mean square error among all values tested. 
By no means are these hyperparameters chosen in the most optimal way, but are indeed sufficiently optimal to achieve accurate predictions, see results in Section \ref{solving class and reg}.
We clarify the specific structure of the hidden layers of the CNN below:
\begin{itemize}
\item the first layer, with 32 kernels;
\item max pooling layer;
\item a dropout layer, with rate $=0.25$;
\item the second layer, with 64 kernels;
\item max pooling layer;
\item a dropout layer, with rate $=0.25$;
\item the third layer, with 128 kernels;
\item max pooling layer;
\item a dropout layer, with rate $=0.4$;
\item a dense layer with 128 units;
\item a dropout layer, with rate $=0.3$.
\end{itemize}

The reason for choosing this structure for our CNN, which is fairly standard for image processing in the computer science discipline, is two fold.
The first is that, heuristically, images are processed by considering the values of each entry of an image matrix together with the neighbouring entries and putting more emphasis on those neighbouring values than entries far from the entry being considered. To study the H\"older regularity of a sample path of a stochastic process, the values of neighbouring points of each entry in a sample path vector will provide the most information about the H\"older regularity of that process; for this reason we employ an image processing-type architecture.
The second is to avoid the task of choosing optimal hyperparameters for the number of filters in each layer. 

If the CNN is able to accurately learn the value of the H\"older exponent, then we are making an important contribution to the field of mathematical finance. 
Indeed, existing calibration techniques for the rBergomi model are still in need of development. 
One calibration technique, suggested by Al\`os and Shiraya \cite{AS19}, is based on their result that the difference between the at-the-money implied volatility for a European Call and the price of a volatility swap has power law behaviour of order $2H$ for short maturities, when the volatility process is driven by a fBm $W^H$. 
The method predicts $H$ accurately on simulated data, but in practice volatility swaps tend to be illiquid for maturities less than 8 months, and therefore it is difficult to use this method for accurate calibration in practice.
Another technique, proposed by Chang \cite{Cha14}, suggests using maximum likelihood estimation to estimate $H$. While the method accurately predicts $H$ from simulated fBm data, the computational cost of this approach is too high for practical application in the quantitative finance industry.
Lastly we consider the least squares method of Gatheral, Jaisson, and Rosenbaum \cite{GJR18}. Inspired by the $q^{th}$ moment formula\footnote{ 
For a fBm $W^H$, the following holds for all $q>0$: 
$E[|W^H_{t+\Delta}-W^H_t|^q] =K_q \Delta^{qH}$, 
where $K_q$ is the absolute $q^{th}$ moment of a standard normal distribution.}
for increments of a fBm, the authors suggest estimation of $H$ via a linear regression of the log of lagged $q^{th}$ moments of the log volatility process against the log of the lags. 
The method, however, is sensitive to the choice of $q$; in particular the method does not perform well for higher order moments. 
Additonally, in Section \ref{solving class and reg}, we show that the least squares method yields erroneous $H$ estimates for processes exhibiting mean reversion.

\section{Solving the regression problem}\label{solving class and reg}
We now move onto solving the regression problem, in order to find the H\"older exponent from sample path input data. 
In Subsection \ref{rBergomi regression} we use the rBergomi model described above to generate our input data, using Cholesky decomposition,
which is the main focus of the paper. 
In Subsection \ref{robustness test regression} we additionally train the CNN with the above rBergomi input data, except with $ \eta \neq 1$, as well as with random $\eta$ and $H$ for each sample path.
We also train with fBm sample paths as input data.
The aim is to illustrate the robustness of this novel method using CNNs.
In Subsection \ref{learning eta} we briefly investigate if the CNN approach can be extended to additionally learn the parameter $\eta$, as well the parameter $H$.

\subsection{rBergomi Model}\label{rBergomi regression}
The sizes of each training/test/validation set of the rBergomi data
are given in Table \ref{rBergomi data set sizes}.
We train the CNN three times: for discretely sampled $H$, for Uniformly sampled $H$, and for Beta sampled $H$.

The CNN calibration method should also be robust to the dimensions of the input data, and training the CNN on vectors of length 100 should produce similar predictive performance to the CNN trained on vectors of length 500.
Consequently we train the CNN with the length of the input vector taking values in 
$\{100,200,300,400,500 \}$.

We present the test results for the CNN in Tables \ref{CNN reg: disc H}, \ref{CNN reg: Unif H}, and \ref{CNN reg: Beta H}. 
We use the mean square error as the loss function in the CNN, and report the predictive performance of the CNN using the root mean square error (RMSE), so that the predicted value and true value of $H$ are of the same unit of measurement.
We also give the time taken, in seconds, to complete the training and testing of the CNN\footnote{All computations were executed in Python, using the Keras module to build and train the network, on a Macbook pro with a 2.6 GHz Intel Core i5 processor and 8 GB 1600 MHz DDR3 memory. The code was run on Google Colaboratory, using the platform's GPU.}.
The Python code is available \href{https://github.com/henrymstone/rough-calibration-using-CNNS}{here}:
https://github.com/henrymstone/CNN-repository.

We train the network with the architecture described above, setting \texttt{batch size}$=64$ as is fairly standard, and \texttt{epochs}$=30$ as this value gave the lowest mean square error.
As a comparison, we also use the least square (LS) calibration approach suggested by \cite[Section 2.1]{GJR18} on the test set used for the CNN, and compute the loss as the root mean square error between the predicted and true values for $H$. 

\begin{table}[h!]
\centering
\begin{tabular}{|c|c|c|c|c|c|}
\hline
Input length & RMSE (CNN) & Training Time (seconds) & Test Time (seconds) & RMSE (LS) & Time (seconds) \\
\hline
100 & $1.041 \times 10^{-2}$ &   69.77 &  0.76   & $2.118 \times 10^{-1}$ &  591.76 \\
\hline
200 & $8.196 \times 10^{-3}$ &   74.89 &  0.75  &  $2.046 \times 10^{-1}$ &  622.00   \\
\hline
300 & $1.096 \times 10^{-2}$ &   80.92&  0.79  & $ 2.025 \times 10^{-1}$ & 635.65  \\
\hline
400 & $ 8.263 \times 10^{-3}$ &  92.02 &  0.93  & $2.014 \times 10^{-1}$ &   634.22 \\
\hline
500 &  $1.232 \times 10^{-2}$ &  93.76 &  0.93   & $ 2.010\times 10^{-1}$ &  627.62 \\
\hline
\end{tabular}
\caption{Test results for discretised $H$.}
\label{CNN reg: disc H}
\end{table}

\begin{table}[h!]
\centering
\begin{tabular}{|c|c|c|c|c|c|}
\hline
Input length & RMSE (CNN) & Training Time (seconds) & Test Time (seconds) & RMSE (LS) & Time (seconds) \\
\hline
100 & $1.137 \times 10^{-2}$ &   66.66 & 0.68   & $1.989 \times 10^{-1}$ &  611.61 \\
\hline
200 & $7.910 \times 10^{-3}$ &   72.20 &   0.73  &  $ 1.927 \times 10^{-1}$ &   620.72 \\
\hline
300 & $5.115 \times 10^{-3}$ &   79.80 &   0.78  & $ 1.907 \times 10^{-1}$ &   630.75 \\
\hline
400 & $9.409  \times 10^{-3}$ &  86.35  & 0.82  & $ 1.895 \times 10^{-1}$ & 634.24  \\
\hline
500 &  $1.282 \times 10^{-2}$ &  96.79 & 0.93  & $ 1.892 \times 10^{-1}$ &  628.26 \\
\hline
\end{tabular}
\caption{Test results for  $H\sim \text{Uniform}(0.0,0.5)$.}
\label{CNN reg: Unif H}
\end{table}

\begin{table}[h!]
\centering
\begin{tabular}{|c|c|c|c|c|c|}
\hline
Input length & RMSE (CNN) & Training Time (seconds) & Test Time (seconds) & RMSE (LS) & Time (seconds) \\
\hline
100 & $6.672 \times 10^{-3}$ &   70.71 & 0.72   & $1.040 \times 10^{-1}$ &  616.74  \\
\hline
200 & $7.193 \times 10^{-3}$ &   73.26  & 0.70   &  $9.962 \times 10^{-2}$ &  626.77  \\
\hline
300 & $1.207 \times 10^{-1}$ &   80.63 & 0.74  & $ 9.791 \times 10^{-2}$ &  637.45 \\
\hline
400 & $1.171  \times 10^{-2}$ &  87.09 & 0.75  & $9.699 \times 10^{-2}$ &   637.33  \\
\hline
500 &  $1.207 \times 10^{-1}$ &  94.20 &   0.77  & $9.663 \times 10^{-2}$ & 644.60 \\
\hline
\end{tabular}
\caption{Test results for  $H\sim \text{Beta}(1,9)$.}
\label{CNN reg: Beta H}
\end{table}

For each input length, the predictive performances of the CNNs trained on  $H\sim\text{Uniform}(0.0,0.5)$ and discretised $H$  are similar, and in each case the CNN approach clearly outperforms the least square approach in terms of predictive power, by one or two orders of magnitude. 
For $H\sim\text{Beta}(1,9)$, the CNN also outperforms the least square approach, again by one or two orders of magnitude, when the input vector length is 100, 200 or 400; the accuracy of the CNN is slightly poorer than the least square approach for the other input vector lengths.

As one would expect for both calibration approaches, the time taken in both cases is, in general,  an increasing function of the length of the input vector for each method of sampling $H$.
Since we are able to train the CNN using Google Colaboratory's GPUs, and Python's Keras module has been optimised for 
execution on GPU, the time taken for training and testing is approximately eight times less than the least square approach.
The training and test times for each sampling method are all very similar.

The above analysis indicates that decreasing the length of the input vector does not significantly worsen the predictive performance of the CNN; in fact when $H\sim\text{Beta}(1,9)$ the performance is improved, when comparing the performance of the length 100 input vector and the length 500 input vector.

\subsection{Robustness Test}\label{robustness test regression}
Following the analysis in Subsection \ref{rBergomi regression}, we set the input vector length to be 100. We then generate our input data by letting $\eta$ take values in $\{0.25, 0.8, 1.3, 2.5\} $ in the rBergomi model, and use discretised sampling for $H$ to generate 5,000 sample paths for each $H$.
We give the results in Table \ref{rBerg regression results different eta}; as before we also include the root mean square error (RMSE) and time taken for the least square (LS) approach of  \cite{GJR18}, applied to the test set as a comparison. Plots of the training error and validation error are given in Appendix \ref{rBerg diff eta loss plot}.

\begin{table}[h!]
\centering
\begin{tabular}{|c|c|c|c|c|c|}
\hline
$\eta$ & RMSE (CNN) & Training Time (seconds) & Test Time (seconds) & RMSE (LS) & Time (seconds) \\
\hline
0.25 & $8.206 \times 10^{-3}$ & 66.81  &   0.63    & $ 2.122 \times 10^{-1}$ & 666.79   \\ 
\hline
0.8 & $1.137 \times 10^{-2}$ &  66.93 & 0.60    & $ 2.122 \times 10^{-1}$ &   665.68    \\ 
\hline
1.3 & $1.473 \times 10^{-2}$ &  67.22 & 0.66    & $2.122 \times 10^{-1}$ &  671.11  \\ 
\hline
2.5 & $9.003 \times 10^{-3}$ &  67.27 & 0.63    & $2.122 \times 10^{-1}$ &  667.16  \\ 
\hline 
\end{tabular}
\caption{rBergomi regression results for $\eta \neq 1$ and input vector length=100.}\label{rBerg regression results different eta}
\end{table}

We can see that the CNN approach vastly outperforms the least square approach in each case, both in terms of the accuracy of the predictions and the time taken. 
The CNN's performance for $\eta=0.8, 1.3$ is slightly worse than for the other two values of $\eta$, but is still superior to the least square approach.
Note that values for the root mean square error for the least square approach are only equal when rounded to three decimal places.

We further extend the robustness test on rBergomi data as follows: we begin by generating 
25,000 $\eta~\sim~\text{Uniform}(0,3)$ and $H\sim\text{Beta}(1,9)$, and then use these values to simulate 25,000 rBergomi sample paths of length 100, each with its own unique and random $\eta$ and $H$. The corresponding training, test, and validation sets thus remain the same size.
The results are presented in Table \ref{rBerg regression: random H and eta}; as above we include the root mean square error and time taken for the least square approach as a comparison. 
Plots of the training error and validation error are given in Appendix \ref{rBerg rand h and Eta}.

\begin{table}[h!]
\centering
\begin{tabular}{|c|c|c|c|c|c|}
\hline
RMSE (CNN) & Training Time (seconds) & Test Time (seconds) & RMSE (LS) & Time (seconds) \\
\hline
$1.382\times 10^{-2}$ & 66.52 &  0.61 & $1.499\times 10^{-1}$ &  598.51 \\ 
\hline
\end{tabular}
\caption{rBergomi regression results for $\eta \sim \text{Uniform}(0,3)$, $H\sim\text{Beta}(1,9)$, and input vector length=100.}\label{rBerg regression: random H and eta}
\end{table}

Interestingly, when $\eta~\sim~\text{Uniform}(0,3)$ and $H\sim\text{Beta}(1,9)$ for each sample path, the predictive power of the CNN is only inferior, in the worst case, by one order of magnitude compared to when the values for either $\eta$ or $H$ are fixed.
Furthermore, the  CNN approach maintains its significant advantage over the least square approach, both in terms of predictive power and speed.

We conclude this robustness test by using fBm sample paths, generated using Cholesky decomposition, as input data to train and test the CNN. For each $H$ we simulate 5,000 sample paths of length 100; we employ the discretised $H$, $H\sim \text{Uniform}(0.0,0.5) $, and  $H \sim \text{Beta}(1,9) $ sampling methods.
The results are given in Table \ref{fBm regression results} and the plots of the training error and validation error are given in Appendix \ref{fBm loss plots}.

\begin{table}[h!]
\begin{tabular}{|c|c|c|c|c|c|}
\hline
Sampling & RMSE (CNN) & Training Time (seconds) & Test Time (seconds) & RMSE (LS) & Time (seconds) \\
\hline
Discretised & $2.483 \times 10^{-2}$& 72.79 &  0.65   & $2.346\times 10^{-1}$ & 635.18  \\ 
\hline 
$\text{Uniform} $ & $2.001 \times 10^{-2}$ & 71.95 &   0.64  & $2.090\times 10^{-1}$ & 615.74  \\ 
\hline
$ \text{Beta}$ & $1.945 \times 10^{-2}$ & 72.20 &  0.67  & $ 9.785  \times 10^{-2}$ & 622.36  \\ 
\hline
\end{tabular}
\caption{fBm regression results for discretised $H$, $H\sim \text{Uniform}(0.0,0.5) $, and  $H \sim \text{Beta}(1,9) $. }\label{fBm regression results}
\end{table}

The CNN maintains its speed advantage over the least square method for each sampling method for $H$, as well as maintaining superior predictive performance, by an order of magnitude.
The results from this final part of the robustness test allow us to conclude that CNNs can indeed identify H\"older regularity from a set of sample paths, thus answering the question posed in the introduction of the paper.

\subsection{Extension to Learning $\eta$}\label{learning eta}
In this Subsection we briefly explore whether the CNN can learn the value of $\eta$, as well as $H$.
We use the rBergomi model, with $\eta \sim \text{Uniform}(0,3)$, $H\sim\text{Beta}(1,9)$ for each sample path, and input vectors of length 100 as our input data, as above; the corresponding output variable then becomes a two-dimensional vector $\mathbf{y}_i=(H_i, \eta_i) $.
We continue to use the root mean square error as a measure of predictive power, and compare to the least square approach as above. The least square (LS) approach can indeed be used to estimate the value of $\eta$ \cite[Section 3.4]{GJR18}, although the authors use the notation $\nu$ instead of $\eta$.
The results are given in Table \ref{table: learning eta}, with loss plots given in Appendix \ref{learning eta loss plots}.

\begin{table}[h!]
\centering
\begin{tabular}{|c|c|c|c|c|c|}
\hline
RMSE (CNN) & Training Time (seconds) & Test Time (seconds) & RMSE (LS) & Time (seconds) \\
\hline
0.666 &  71.82 & 0.62 & 1.170 & 613.62 \\ 
\hline
\end{tabular}
\caption{Regression results for learning $H$ and $\eta$, with input vector length=100.}\label{table: learning eta}
\end{table}

Encouragingly, we see that the CNN approach still outperforms the least square approach for both accuracy and time; however the CNN approach is approximately twice as accurate as the least square approach, compared to the orders of magnitude in the cases above.
Note that we kept all hyperparameter values unchanged; it is likely that superior predictive power could be achieved with some tuning of the hyperparameter values in the CNN. This is not the aim of this paper, however, and we leave this once again to further research.

\begin{remark}
We conclude the above results with a remark on the speed and accuracy advantages of the CNN approach, from a theoretical perspective. 
The above results show the CNN method to be more accurate, by orders of magnitude, and significantly faster than the existing method suggested in \cite{GJR18} when estimating $H$ on simulated rBergomi and fBm data. 
The explanation for the speed advantage of the CNN method is that once the network has been trained, estimations are made by the straightforward computation of the composition \eqref{eqn: NN} given in Definition \ref{neural net def}; 
on the other hand, the LS approach requires a number of successive regressions to be executed in order to estimate $H$. Clearly then the CNN method will be faster than the LS approach.
We now recall our motivation that justifies the use of a CNN to estimate $H$: the predictive power of the CNN lies in the convolutional operator, which assigns a value to each entry $\mathbf{x}_i^j $ of an input vector $\mathbf{x}_i$; this value is determined by the relative values of the entries  neighbouring $\mathbf{x}_i^j$. 
To estimate the H\"older regularity of a stochastic process, the values of neighbouring points of each entry in a trajectory vector will provide the most information about the H\"older regularity of that trajectory, and thus a CNN is indeed a valid means of estimating $H$.
The CNN is able to detect very subtle path regularity properties, via interated applications of the convolutional operator, allowing for very accurate predictions of $H$. Furthermore, the LS estimation is based on an approximation of the $q^{th}$ moment of an increment of log volatility, whose accuracy is highly dependent on the choice of $q$, while the CNN approach is completely independent of the choide of $q$.
Together, this explains why the CNN approach gives more accurate $H$ estimations than the LS approach.
\end{remark}

\subsection{\textbf{Robustness test on a (mean-reverting) Ornstein-Uhlenbeck process}}\label{LS mis-estimation}
We finish Section \ref{solving class and reg} with a comparison of the performance of the CNN and LS approaches on a (mean-reverting) Ornstein-Uhlenbeck process.
Recall that a (mean-reverting) Ornstein-Uhlenbeck process $X$ satisfies the following SDE:
$$ \D X_t = (a-b X_t)\D t + c X_t \D W_t, \qquad X_0 = x_0 \in \mathbb{R},
$$
where $W$ is a standard Brownian motion, and that $X$ is $\gamma$-H\"older continuous for all $\gamma \in (0,1/2)$; therefore we expect both approaches to estimate $H\approx 0.5$.

We set the input length to be 100 for the CNN, using the Discretised and Uniform $H$ sampling methods detailed above\footnote{We do not experiment with the Beta sampling method, since those $H$ values are concentrated around 0.1, and so will produce poor estimations when we expect $H\approx 0.5.$}. Each trained network is then used to estimate $H$ on trajectories of a (mean-reverting) Ornstein-Uhlenbeck process, which are simulated using the Euler-Maruyama method. 
We simulate 1000 trajectories of length 100, setting $(x_0,a,b)=(0.1, 1., 2.1)$, and report the mean estimated $H$ values for the CNN and LS approaches in Table \ref{mean rev table} below, for the `high volatity' ($c=3$) and `low volatility' ($c=0.3$) regimes.

\begin{table}[h!]
\centering
\begin{tabular}{|c|c|c|c|}
\hline
$c$ & CNN (Discretised) & CNN (Uniform)  & LS \\
\hline
3 &  0.46 &  0.42 & 0.22 \\ 
\hline
0.3 &  0.49 & 0.44 & 0.67 \\
\hline 
\end{tabular}
\caption{Mean $H$ estimates for a (mean-reverting) Ornstein-Uhlenbeck process $X$.}\label{mean rev table}
\end{table}

The results in Table \ref{mean rev table} provide compelling evidence that the CNN is indeed learning the H\"older regularity of the sample path, as the mean $H$ estimates for the CNN are sufficiently close to 0.5. 
Note that the mean LS $H$ estimates are significantly further from 0.5.
The results provide further evidence that the CNN approach is preferable to the LS approach, which is not only slower and less accurate than the CNN approach but may also incorrectly identify roughness in data.

\section{Calibration using CNNs}\label{calibration section}

We begin by using the trained CNNs from Section \ref{solving class and reg} to predict the H\"older exponent of historic realised volatility data from the 
\href{https://realized.oxford-man.ox.ac.uk/data/download}{Oxford-Man Institute of Quantitative Finance}, 
which is free and publicly available\footnote{https://realized.oxford-man.ox.ac.uk/data/download}.
We choose the length of the input vectors to be 100 from the analysis given in Section \ref{solving class and reg}.
We took a sample of 10 different indices\footnote{AEX, All Ordinary, DAX, FTSE 100, Hang Seng, NIFTY 50, Nasdaq 100,	Nikkei 225, S\&P500 , Shanghai Composite.} from the 31 available; for each index we then used a time series of 200 sequential data points to create 11 vectors of length 100  (entries 0 to 100, 10 to 110, and so on) to predict 
the H\"older exponent for each index. 
We compute the root mean square error between the CNN prediction and the least square prediction, and the standard deviation of the difference between the two predictions; see Table \ref{calibration results on real data}.

\begin{table}[h!]
\centering
\begin{tabular}{|c|c|c|}
\hline
Sampling Method & Root Mean Square Error & Standard Deviation \\
\hline
 Discretised $H$ & $5.558 \times 10^{-2}$ & $2.900\times 10^{-3}$ \\
 \hline
 $H\sim \text{Uniform}(0.0,0.5) $ &  $ 2.444 \times 10^{-1}$ & $1.141 \times 10^{-2}$ \\
\hline
 $H\sim \text{Beta}(1,9)$ & $4.253 \times 10^{-2}$ & $1.098 \times 10^{-3}$ \\
\hline
\end{tabular}
\caption{Calibration results}\label{calibration results on real data}
\end{table}

Indeed, we can see that this set of results is very promising. In each case, both the root mean square error and the standard deviation are small; note that both root mean squared error values and the standard deviations of discretised $H$ and $H\sim\text{Beta}(1,9)$ are an order of magnitude greater than for $H\sim\text{Uniform}(0.0,0.5)$. 
This therefore indicates that the calibration values attained by the network are very close to those attained by the least squares approach. Note that this in turn provides further evidence that $H\approx 0.1$, further corroborating the findings of \cite{GJR18} and  \cite{BLP16}.

We can state, therefore, that this calibration scheme is precise enough to be used in practice, where we recommend using $H\sim\text{Beta}(1,9)$ to train the network with input length 100, due smaller root mean squared error and standard deviation in testing, the nondiscretised network output, and  emphasis of this distribution on ``rough'' values of $H$ i.e. $H\approx 0.1$.

The practical implementation of our calibration scheme is a simple two-step process. 
The first step is to train the CNN, with the $H$ sampling method and input length $n$  chosen by the practitioner; this can be done once offline, with the weights of the trained network saved to avoid unnecessary repetition for each calibration task. 
The second step is to input the most recent $n$ volatility observations  into the CNN, which will return the corresponding $H$ value for those $n$ observations. This value of $H=\alpha + 1/2$ can then be inputed into the rBergomi model and used for, say, pricing. 
We note that some testing for the optimal choice of $H$ sampling method and input length $n$ is required on the part of the practitioner implementing our calibration scheme.

\begin{remark}\label{sup/unsup remark}
We now take this opportunity to discuss the calibration methodology presented in this paper.
We are treating calibration as a \textit{supervised} learning problem when in practice it is an \textit{unsupervised} learning problem, strictly speaking.
While each vector in the input data in the regression problem given in Section \ref{solving class and reg} was indeed labelled with the corresponding $H$ value, the data from the Oxford-Man Institute `realized' library has no such labels, thus we use the least square calibration values as ``true'' values.
\end{remark}

\section{discussion of results and conclusion}\label{conclusion}
The results presented in the paper are very promising: CNNs can indeed be used to give highly accurate predictions for the H\"older exponent of rBergomi and fBm sample paths. In fact, we conjecture that even higher accuracy could be attained if we optimised the hyperparameter values in our CNN. Additionally, invoking a k-fold cross validation could further improve the predictive power of the CNN. These two techniques could additionally be used to attain more accurate predictions for the parameter $\eta$ in the rBergomi model.

The results in the paper invite a number of intriguing questions for future research. 
Firstly, is it possible to use the unlabelled volatility data, together with known forecasting formula, to train the network and thus predict ``roughness'', potentially improving this calibration method further and addressing Remark \ref{sup/unsup remark}?
Secondly, can we achieve more accurate predictions in the case where the input data are fBm sample paths or when we are also trying to learn the value of $\eta$?
Lastly, we propose the following question: is it possible to achieve similar predictive results using more traditional machine learning techniques, such as ensemble and boosting methods?

To conclude, in this paper we have shown for the first time that CNNs can indeed learn the ``roughness'' 
(i.e. the H\"older exponent) of time series data with a high degree of accuracy. 
We have additionally provided an efficient and accurate means of calibrating the rBergomi model.
We have also shown our method to be far more accurate, by orders of magnitude, and significantly faster than the existing method suggested in \cite{GJR18}.
We have show our method to correctly estimate $H$ on mean-reverting trajectories, where the least squares approach fails and incorrectly identifies roughness.
Finally, the paper has also opened up a number of interesting avenues for future research.

\appendix
\section{Loss plots for rBergomi with different $H$ sampling}\label{Loss plots}
\begin{figure}[h!]
\includegraphics[scale=0.4]{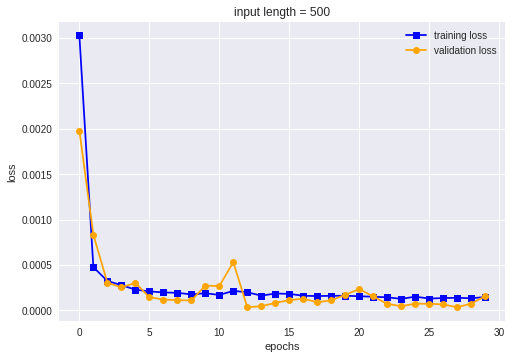}
\hspace{3mm}
\includegraphics[scale=0.4]{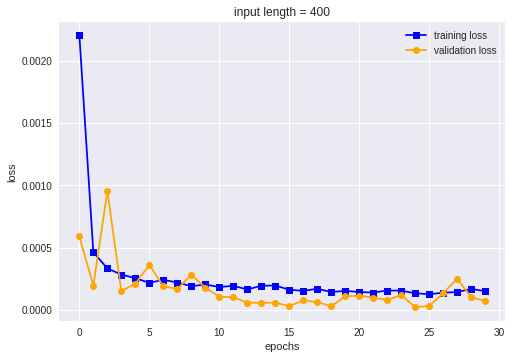}
\hspace{3mm}
\includegraphics[scale=0.4]{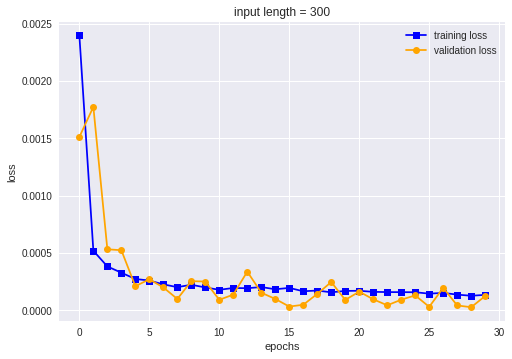}
\hspace{3mm}
\includegraphics[scale=0.4]{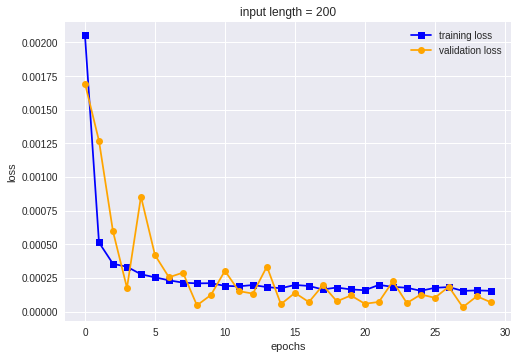}
\hspace{3mm}
\includegraphics[scale=0.4]{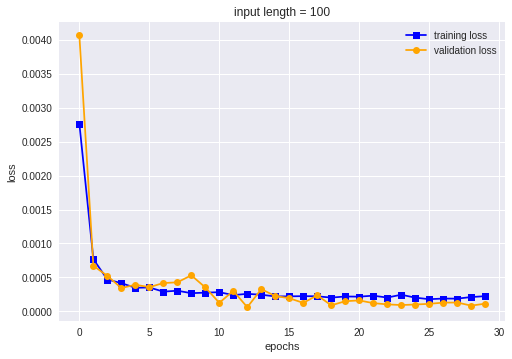}
\caption{Loss plots for discretised $H$.}\label{loss plot disc H} 
\end{figure}

\begin{figure}[h!]
\includegraphics[scale=0.4]{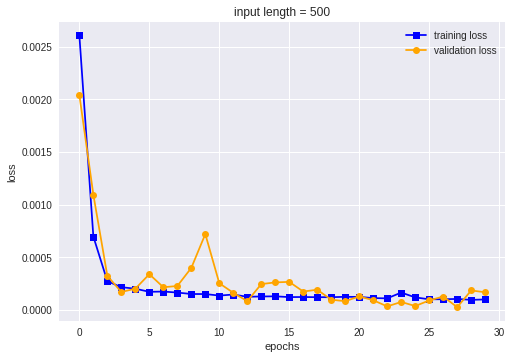}
\hspace{3mm}
\includegraphics[scale=0.4]{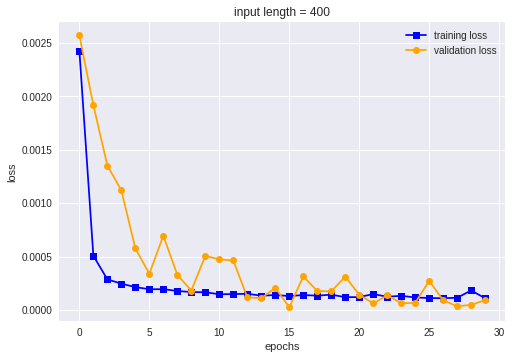}
\hspace{3mm}
\includegraphics[scale=0.4]{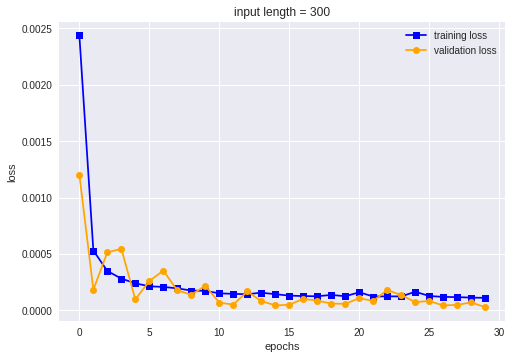}
\hspace{3mm}
\includegraphics[scale=0.4]{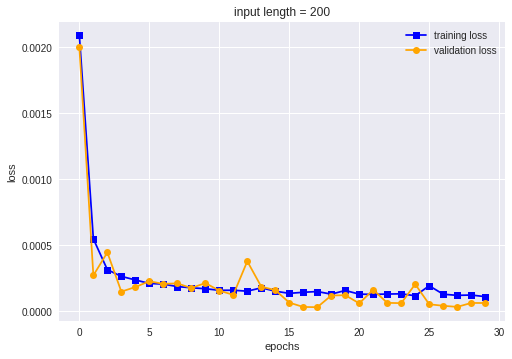}
\hspace{3mm}
\includegraphics[scale=0.4]{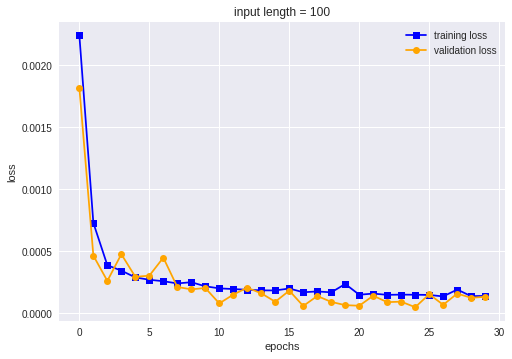}
\caption{Loss plots for $H\sim \text{Uniform}(0.0,0.5)$.}\label{loss plot unif H} 
\end{figure}

\begin{figure}[h!]
\includegraphics[scale=0.4]{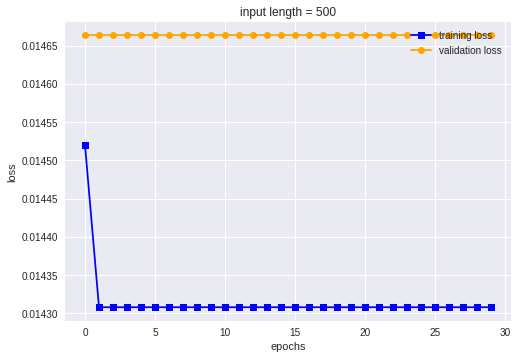}
\hspace{3mm}
\includegraphics[scale=0.4]{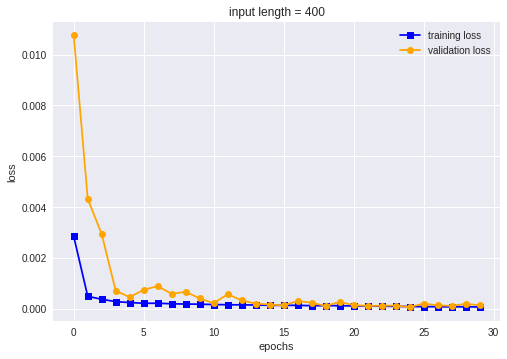}
\hspace{3mm}
\includegraphics[scale=0.4]{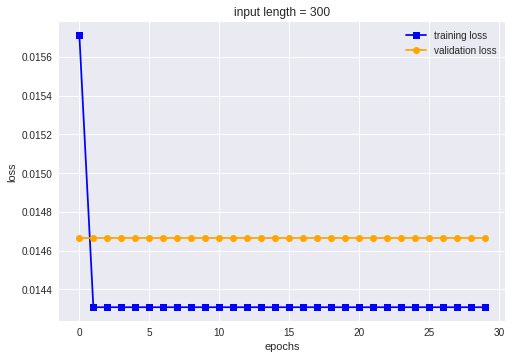}
\hspace{3mm}
\includegraphics[scale=0.4]{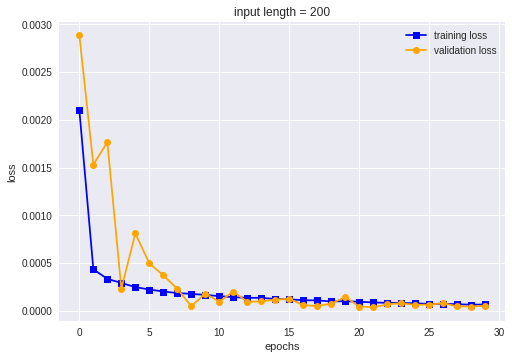}
\hspace{3mm}
\includegraphics[scale=0.4]{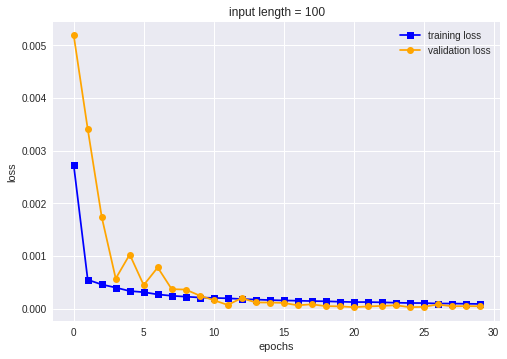}
\caption{Loss plots for $H\sim\text{Beta}(1,9)$.}\label{loss plot beta H} 
\end{figure}

Here we plot the training loss and validation loss (MSE) at each epoch, for discretised $H$, $H\sim \text{Uniform}(0.0,0.5) $, and  $H \sim \text{Beta}(1,9) $ in
Figures  \ref{loss plot disc H}, \ref{loss plot unif H}, and \ref{loss plot beta H} respectively.
Note that the training loss and validation loss both tend to decrease as the number of epochs increases; this is a good indication that predictive performance improves as the number of epochs increases, without overfitting \cite[Overfitting Section, Page 619]{KNTY18}. 
The only exception is  Figure \ref{loss plot beta H}, in the case where input length is either 300 or 500; note that these two cases also correspond to larger root mean squared error values.
Recall, however, that our calibration scheme uses input vectors of length 100 so this does not pose any real problems for practical use.

\vspace{2mm}

\clearpage

\section{Loss plots for rBergomi with $\eta \neq 1$}\label{rBerg diff eta loss plot}
\begin{figure}[h!]
\includegraphics[scale=0.4]{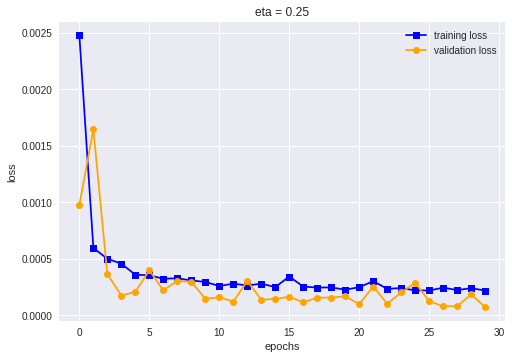}
\hspace{3mm}
\includegraphics[scale=0.4]{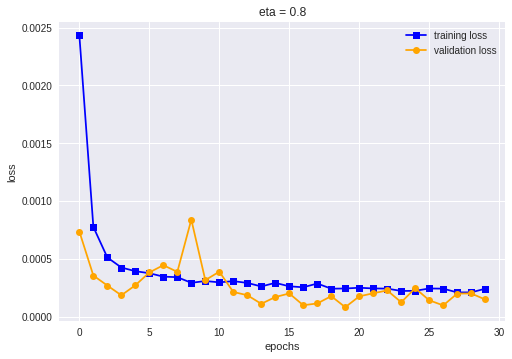}
\hspace{3mm}
\includegraphics[scale=0.4]{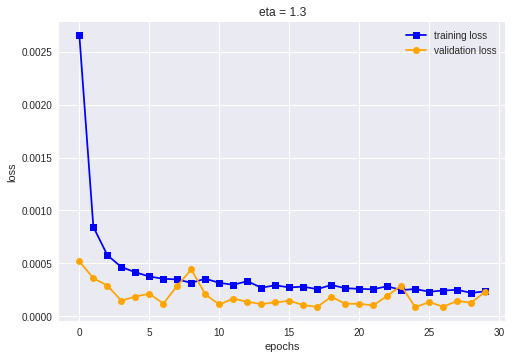}
\hspace{3mm}
\includegraphics[scale=0.4]{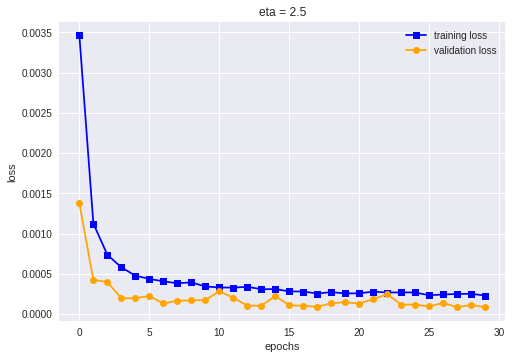}
\caption{Loss plots for discretised $H$ and $\eta\in \{0.25,0.8,1.3,2.5 \}$.}\label{loss plot rBerg diff eta} 
\end{figure}

We now plot the training loss and validation loss (MSE) at each epoch, as above, for the rBergomi model with discretised $H$ and $\eta\in \{0.25,0.8,1.3,2.5 \}$ in Figure \ref{loss plot rBerg diff eta}. We fix the input vector length to be 100.
For each $\eta$ value, the training loss and validation loss both decrease with each epoch.

\clearpage

\section{Loss plots for rBergomi with $\eta\sim\text{Uniform}(0,3)$ and $H\sim\text{Beta}(1,9)$}\label{rBerg rand h and Eta}
\begin{figure}[h!]
\includegraphics[scale=0.5]{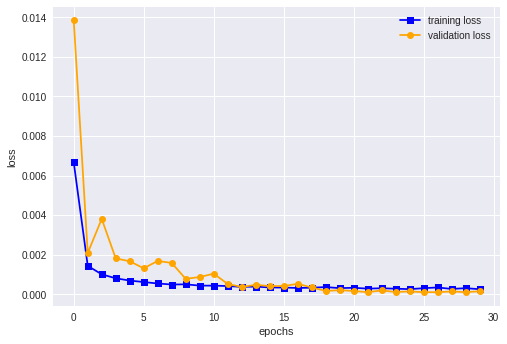}
\caption{Loss plots for $\eta\sim\text{Uniform}(0,3)$ and $H\sim\text{Beta}(1,9)$.}\label{loss plot rBerg random H and eta} 
\end{figure}

We now plot the  training loss and validation loss (MSE) at each epoch, as above, for the rBergomi model with 
$\eta\sim\text{Uniform}(0,3)$ and $H\sim\text{Beta}(1,9)$ for each sample path, in Figure \ref{loss plot rBerg random H and eta}. We fix the input vector length to be 100.
Note that the training loss and validation loss both decrease with each epoch.

\clearpage

\section{Loss plots for fBm with different $H$ sampling}\label{fBm loss plots}
We plot the training and validation loss (MSE) at each epoch, for fBm with discretised $H$, $H~\sim~\text{Uniform}(0.0,0.5) $, and  $H \sim \text{Beta}(1,9) $ in
Figures  \ref{loss plot fBm disc H}, \ref{loss plot fBm unif H}, and \ref{loss plot fBm beta H} respectively. As in Appendix \ref{rBerg diff eta loss plot} we fix the input vector length to be 100.
For discretised $H$ and $H\sim \text{Uniform}(0.0,0.5) $ the training loss and validation loss both decrease with each epoch. For $H \sim \text{Beta}(1,9) $, however, the training loss decreases but the validation loss remains flat. This could possibly suggest overfitting, and poor predictive performance; note that the root mean square error was indeed higher than for discretised $H$ and $H\sim \text{Uniform}(0.0,0.5) $.

\begin{figure}[h!]
\centering
\includegraphics[scale=0.4]{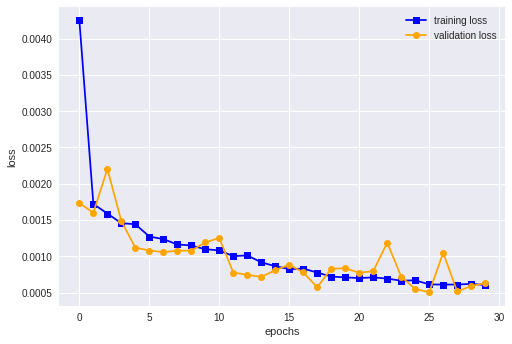}
\caption{Loss plot for fBm with discretised $H$.}\label{loss plot fBm disc H}
\end{figure}

\begin{figure}[h!]
\centering
\includegraphics[scale=0.4]{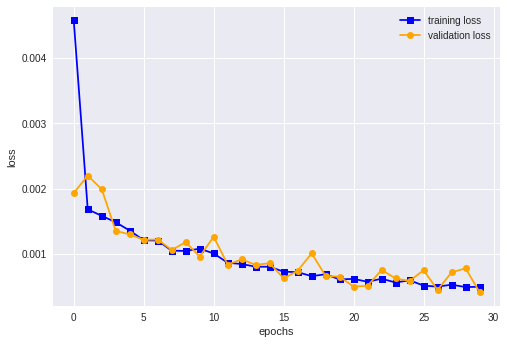}
\caption{Loss plot for fBm with $H\sim\text{Uniform}(0.0,0.5)$. }\label{loss plot fBm unif H}
\end{figure}

\begin{figure}[h!]
\centering
\includegraphics[scale=0.4]{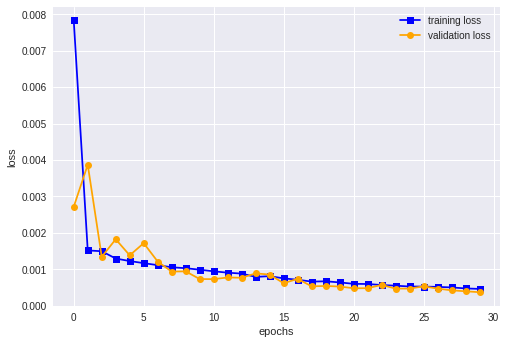}
\caption{Loss plot for fBm with $H\sim\text{Beta}(1,9)$. }\label{loss plot fBm beta H}
\end{figure}

\clearpage

\section{Loss plots for learning $\eta$}\label{learning eta loss plots}

In Figure \ref{plot: learning eta plot} we plot the training loss and validation loss (MSE) at each epoch, as in Appendix \ref{rBerg rand h and Eta}, for the rBergomi model with 
$\eta\sim\text{Uniform}(0,3)$ and $H\sim\text{Beta}(1,9)$ for each sample path. Recall that in this case the CNN is learning the value of $\eta$, as well as the value of $H$.  
We fix the input vector length to be 100.
Note that the training loss and validation loss both tend to decrease with each epoch.

\begin{figure}[h!]
\includegraphics[scale=0.5]{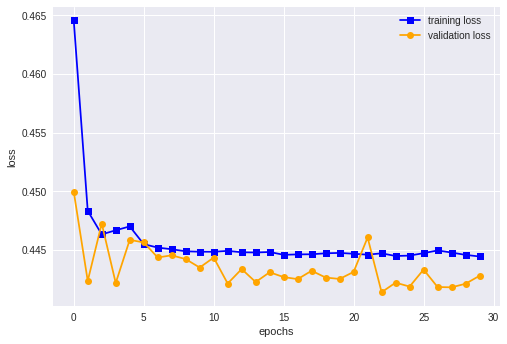}
\caption{Loss plots for $\eta\sim\text{Uniform}(0,3)$ and $H\sim\text{Beta}(1,9)$.}\label{plot: learning eta plot}
\end{figure}

\end{document}